\documentclass[twocolumn,aps,prl,showpacs,groupedaddress,superscriptaddress
]{revtex4-1}

\usepackage[utf8]{inputenc}  
\usepackage[T1]{fontenc}

\usepackage[normalem]{ulem}

\usepackage[dvipsnames]{xcolor}
\usepackage{graphicx,epsfig}
\usepackage{braket}
\usepackage{color}
\usepackage{amsmath}
\usepackage{amsfonts}
\usepackage{fancyhdr}
\usepackage{mathrsfs}
\usepackage{mathtools}

\usepackage[pdftex,linkcolor=blue,citecolor=blue,urlcolor=blue,colorlinks]{hyperref}

\newcommand{\mean}[1]{\langle#1\rangle}

\newcommand{\dgr}{^{\dagger}}

\DeclarePairedDelimiter\abs{\lvert}{\rvert}%
\DeclarePairedDelimiter\norm{\lVert}{\rVert}%
\makeatletter
\let\oldabs\abs
\def\abs{\@ifstar{\oldabs}{\oldabs*}}
\let\oldnorm\norm
\def\norm{\@ifstar{\oldnorm}{\oldnorm*}}
\makeatother

\renewcommand{\vec}[1]{\mbox{\boldmath$#1$}}

\newcommand{\enum}[1]{\mathit{e}^{#1}}

\newcommand{\ham}{\hat{H}}
\newcommand{\Spvek}[2][r]{%
  \gdef\@VORNE{1}
  \left(\hskip-\arraycolsep%
    \begin{array}{#1}\vekSp@lten{#2}\end{array}%
  \hskip-\arraycolsep\right)}
\newcommand{\appropto}{\mathrel{\vcenter{
\offinterlineskip\halign{\hfil$##$\cr
  \propto\cr\noalign{\kern2pt}\sim\cr\noalign{\kern-2pt}}}}}

\begin{abstract}
In spinor Bose-Einstein condensates, spin-changing collisions are a remarkable proxy to coherently realize  macroscopic many-body quantum states. These processes have been, e.g., exploited to generate entanglement, to study dynamical quantum phase transitions, and proposed for realizing nematic phases in atomic condensates. In the same systems dressed by Raman beams, the coupling between spin and momentum induces a spin dependence in the scattering processes taking place in the gas. Here we show that, at weak couplings, such modulation of the collisions leads to an effective Hamiltonian which is equivalent to the one of an {\it artificial} spinor gas with spin-changing collisions that are  {\it tunable} with the Raman intensity. By exploiting this dressed-basis description, we propose a robust protocol to coherently drive the spin-orbit coupled condensate into the ferromagnetic stripe phase via crossing a quantum phase transition of the effective low-energy model in an excited-state.
\end{abstract}

\begin{document}
\title{Dynamical preparation of stripe states in spin-orbit coupled gases}
\author{J. Cabedo}
\affiliation{Departament de F\'isica, Universitat Aut\`onoma de Barcelona, E-08193 Bellaterra, Spain.}
\author{J. Claramunt}
\affiliation{Department of Mathematics and Statistics, Lancaster University, Lancaster LA1 4YW, United Kingdom.}
\author{A. Celi}
\affiliation{Departament de F\'isica, Universitat Aut\`onoma de Barcelona, E-08193 Bellaterra, Spain.}


\maketitle

Artificial spin-orbit coupling (SOC) in ultracold atom gases offers an excellent platform for studying quantum many-body physics \cite{Dalibard-2011, Goldman-2014, Maciej-Book-2012}. The interplay between light dressing induced by Raman coupling \cite{Lin-2009} and atom-atom interactions can lead, for instance, to high-order synthetic partial waves \cite{Williams-2012}, to chiral interactions and density-dependent gauge fields \cite{Chiral-interactions-2020} or to the formation of stripe phases \cite{Lin-2011}. The latter have gained significant attention over the last decade \cite{Wang-prl-2010, Ho-prl-2011, Martone-2012, Li-2013, Lan-2014, Hou-2018}, in great part due to its supersolid-like properties \cite{Chester-pra-1970, Leggett-prl-1970, Boninsegni-rmp-2012}, that is, its simultaneous spontaneous breaking of translational invariance and of U(1) (global) phase symmetry, resulting in a crystalline structure that maintains off-diagonal long-range order.

Accessing the stripe regime of ultracold gases with SOC remains experimentally challenging, since its stability relies on the asymmetry between intra- and inter-spin interactions, typically small in common spinor BECs. The predicted spatial density modulations have only been unambiguously observed in \cite{Li-2017}, using orbital states in a superlattice as pseudo-spin states,  and very recently also in metastable states of a $^{87}$Rb spinor gas \cite{Putra-2020} (for its realization in dipolar gases, see \cite{Tanzi-ss-prl-2019, Bottcher-prx-2019, Chomaz-prx-2019}). While sharing many properties with conventional supersolids, the nature of the stripe phases in gases with SOC is still debated \cite{Hofmann-2021}, with current proposals focusing on probing its excitation spectrum. So far, most protocols to enhance the accessibility of the phase and the contrast of the stripes pursue an effective decrease of the intraspin interactions \cite{Li-prl-2016, Geier-arXiv-2021}. Alternatively, here we propose a novel approach to access the stripe regime of a spin-1 gas with largely symmetric spin interactions, based on the coherent spin-mixing dynamics induced by Raman dressing. 

\begin{figure}[b!]
\includegraphics[width=0.99\linewidth]{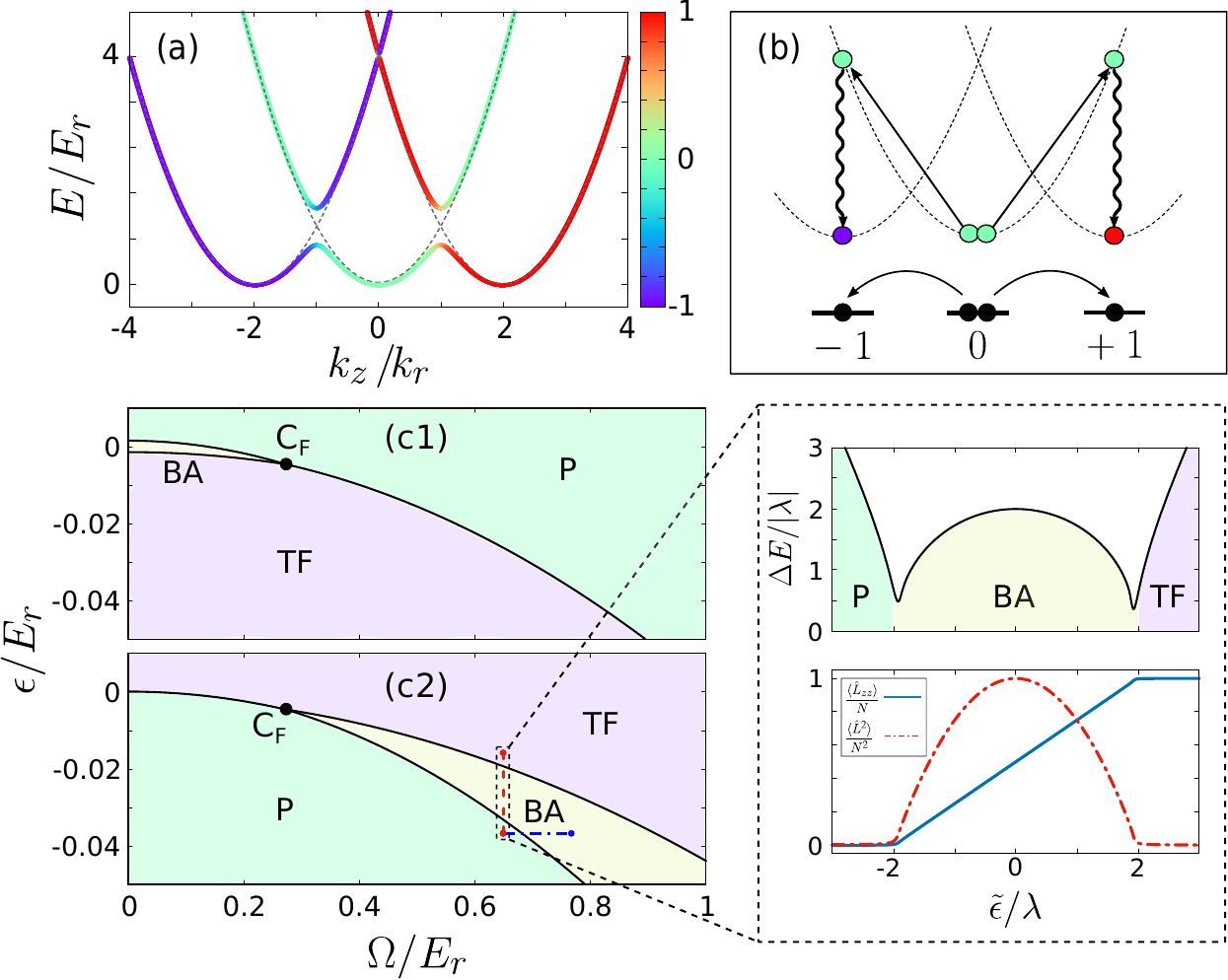}
\caption{(Color online) \textbf{Pseudospin dynamics in SOC BECs.} (a) Dispersion bands of the dressed Hamiltonian $\hat{\mathcal{H}}_{\mathrm{k}}$ with $\Omega= 0.65 E_r$, $\delta=0$ and $\epsilon=\Omega^2/16E_r$. The color texture indicates the expected value of the spin of the  dressed states. Dashed lines show the undressed dispersion bands. (b) Schematic representation of resonant collisions mediated by Raman transitions (represented in wavy lines) which act as effective spin-changing collisions. For weak Raman coupling and interactions, the dressed-state dynamics can be captured by the pseudospin Hamiltonian \eqref{eq_spin_ham_mz0}. (c1) Phase diagram of \eqref{eq_spin_ham_mz0}, as a function of the Raman Rabi frequency $\Omega$ and effective quadratic Zeeman shift $\epsilon$, for $^{87}$Rb at $n=7.5\cdot10^{13}$ cm$^{-3}$. The polar (P), twin-Fock (TF) and broken-axisymmetry (BA) phases meet at the tricritical point $C_F$ (black dot). (c2) Corresponding phase diagram for the highest-excited eigenstate. The upper panel in the inset shows the energy gap between the two most excited eigenstates along the red dashed segment for $N=1000$. The lower panel shows the expected value of the collective pseudospin $\hat{L}^2$ (dashed red) and tensor magnetization $\hat{L}_{zz}$ (solid blue)}.\label{Fig_eff_low_en_phys} 
\end{figure}

Several authors have suggested a connection between spinor gases with spin-changing collisions and SOC BECs \cite{Higbie-2004, Lian-sr-2013, Lan-2014, Huang-sr-2015, Huang-pra-2017, Cabedo-2019, Chen-pra-2020,Lao-prl-2020}. In this work, we show analytically that the Raman-dressed spin-1 SOC gas at low energy is equivalent, for weak Raman coupling and interactions, and zero total magnetization, to an artificial spin-1 gas with {\it tunable} spin-changing collisions. Under these conditions, the system is well described by a one-axis-twisting Hamiltonian \cite{Kitagawa-1993,Law-1998}. Such Hamiltonian explains several quantum many-body phenomena in spinor condensates \cite{Ho-prl-1998, Stamper-Kurn-2013}, including the generation of macroscopic entanglement \cite{Duan-2000, Bookjans-2011, Lucke-science-2011, Gross-nature-2011, Hamley-nature-2012, Zhang-prl-2013, Gabbrielli-prl-2015, Peise-ncomms-2015, Hoang-PNAS-2016,  Luo-2017, Zou-2018, Kunkel-science-2018, Pezze-prl-2019,  Qu-prl-2020}, with potential metrological applications \cite{Pezze-review-2018}, and the observation of nonequilibrium phenomena such as the formation of spin domains and topological defects \cite{Stenger-nature-1998, Sadler-2006, Bookjans-prl-2011, Vinit-2013, Hoang-natcoms-2016, Anquez-prl-2016, Prufer-nature-2018, Chen-prl-2019, Kang-prl-2019, Jimenez-Garcia-2019, Prufer-natphys-2020}. Recently, dynamical \cite{Heyl-rpp-2018} and excited-state \cite{Cejnar-arxiv-2020} quantum phase transitions have been theoretically \cite{Ceren-pra-2018, Feldmann-arxiv-2020-ESQPT} and experimentally \cite{Yang-PRA-2019, Tian-PRL-2020} studied in spin-1 BECs with spin-changing collisions.
Here we exploit this map to provide a many-body protocol to access the ferromagnetic stripe phase of the SOC gas via crossing a quantum phase transition of the low-energy Hamiltonian in an excited state. This preparation enhances the accessibility of the phase, which has as ground-state phase a very narrow region of stability \cite{Martone-2016} and has not been experimentally demonstrated  so far.

\textit{System.---} We consider a spin-1 Raman-dressed Bose gas held in an isotropic harmonic potential $V_\mathrm{t} = \frac{1}{2}m\omega_\mathrm{t}^2 \vec{r}^2$ with the atoms interacting via two-body s-wave collisions. In a frame corotating and comoving with the laser beams, the system is described by the Hamiltonian
$\ham \!\!=\!\! \int \! d \vec{r} \! \left[ \hat{\vec{\psi}}^{\dagger}\!\! \left(\hat{\mathcal{H}}_{\mathrm{k}} \!+\! V_\mathrm{t}\right) \hat{\vec{\psi}} + \frac{g_0}{2}(\hat{\vec{\psi}}^{\dagger}\hat{\vec{\psi}})^2 \!+\! \frac{g_2}{2}\sum_j (\hat{\vec{\psi}}^{\dagger} \hat{F}_j \hat{\vec{\psi}})^2 \right]$, with $\hat{\vec{\psi}} = (\hat{\psi}_{-1},\hat{\psi}_{0},\hat{\psi}_{1})^T$ being the spinor field operator and $\{\hbar\hat{F}_x,\hbar\hat{F}_y, \hbar\hat{F}_z\}$ being the spin-1 matrices. Here $g_0 = 4\pi \hbar^2(a_0+ 2a_2)/3m$ and $g_2 = 4\pi \hbar^2(a_2-a_0)/3m$,  with $a_0$ and $a_2$ being the scattering lengths in the $F=0$ and $F=2$ channels, respectively. The dressed kinetic Hamiltonian reads $\hat{\mathcal{H}}_{\mathrm{k}} = \frac{\hbar^2}{2m}\left(\vec{k} - 2k_r \hat{F}_z \vec{e}_z\right)^2 + \frac{\Omega}{\sqrt{2}}\hat{F}_x + \delta \hat{F}_z + \epsilon \hat{F}_z^2$, where $\Omega$ is the Raman coupling strength, $\delta$ is the Raman detuning and $\epsilon$ is the effective quadrupole tensor field strength. The latter term can be controlled independently of $\delta$ by employing two different Raman couplings between the two Zeeman pairs $\{\ket{1,1},\ket{1,0}\}$ and $\{\ket{1,0},\ket{1,-1}\}$, and simultaneously adjusting the Raman frequency differences \cite{Campbell-2016}. We label the Raman single-photon recoil energy and momentum as $E_r = \frac{\hbar^2 k_r^2}{2m}$ and $\hbar k_r$, respectively. In the weakly-coupled regime, the lowest dispersion band of $\hat{\mathcal{H}}_{\mathrm{k}}$ presents a triple-well shape along the direction of the momentum transfer, which we arbitrarily set along the $\hat{z}$ axis. Spin texture is present in the band, with the spin mixture being the largest at the vicinity of the avoided crossings (see Fig.\,\ref{Fig_eff_low_en_phys}(a)). While much smaller, the spin overlap between states located at the vicinity of adjacent minima is nonzero, and increases linearly with $\Omega$. This overlap allows collision processes that exchange large momentum at low energies. These Raman-mediated processes act as spin-changing collisions, as illustrated in Fig.\,\ref{Fig_eff_low_en_phys}(b).

\textit{Low-energy effective theory.---} We now consider the regime where  $\delta$, $\epsilon$, $\hbar\omega_\mathrm{t}$ and the interaction energy per particle are all much smaller than the recoil energy $E_r$. Such low-energy landscape is well captured by an effective theory in which all the dynamics involves only the lowest band modes around each band minima $\vec{k}_j \sim 2 j k_r \vec{e}_z$, with $j \in \left\{-1,0,1\right\}$. Under these considerations, we re-express the spinor field $\hat{\vec{\psi}}$ in terms of the lowest-band dressed fields at the vicinity of each $\vec{k}_j$, which we label as $\hat \varphi_j$, and set a cut-off $\Lambda \ll \hbar k_r$ to the momentum spread $\vec{p}$ around them. With this notation, we can identify the operators acting in the separated regions as a pseudospinor field $\hat{\vec{\varphi}} =  (\hat \varphi_{-1},\hat \varphi_{0},\hat \varphi_{1})^T$, with $\left[\hat\varphi_{i}(\vec{p}), \hat\varphi_{j}\dgr(\vec{p}')\right] = \delta(\vec{p}-\vec{p}')\delta_{ij}$. By using perturbation theory up to second order in $\Omega$, the low-energy Hamiltonian can be written as $\ham \simeq \ham_\mathrm{S} + \ham_\mathrm{A}$ (see Supplemental Material for more details). Here $\ham_\mathrm{S}$ and $\ham_\mathrm{A}$ include the pseudospin-symmetric and nonsymmetric contributions, respectively, given by
\small
\begin{equation}\label{eq_ham_symmetric}
\ham_\mathrm{S} = \!\int\! d \vec{r} \Bigg[\sum_i  \hat{\varphi}_{i}^{\dagger} \left(\frac{\vec{p}^2}{2m} + V_\mathrm{t}\right) \hat{\varphi}_{i} + \frac{g_0}{2} \sum_{ij}  \hat{\varphi}_{i}^{\dagger}\hat{\varphi}_{j}^{\dagger}\hat{\varphi}_{j}\hat{\varphi}_{i}\Bigg],
\end{equation} 
\normalsize
and
\small
\begin{align}\label{eq_intham_nonsymmetric}
\ham_\mathrm{A} &= \!\int\! d\vec{r}\Bigg[\frac{g_2}{2}\sum_j (\hat{\vec{\varphi}}^{\dagger} \hat{F}_j \hat{\vec{\varphi}})^2  + \tilde{g}_2\left(\hat{\varphi}_{1}^{\dagger}\hat{\varphi}_{1} + \hat{\varphi}_{-1}^{\dagger}\hat{\varphi}_{-1}\right)\hat{\varphi}_{0}^{\dagger}\hat{\varphi}_{0} \cr
+&\tilde{g}_2 \left( \hat{\varphi}_{1}^{\dagger}\hat{\varphi}_{-1}^{\dagger}\hat{\varphi}_{0}\hat{\varphi}_{0} + \text{H.c}  \right) + \vec{\hat{\varphi}}\dgr\left( \delta \hat{F}_z + \tilde{\epsilon} \hat{F}_z^2\right) \vec{\hat{\varphi}} \Bigg],\end{align}
\normalsize
with $\tilde{g}_2 = g_0\frac{\Omega^2}{16 E_r^2}$. The coefficient $\tilde{\epsilon}$ includes the correction to $\epsilon$, with $\tilde{\epsilon} = \epsilon +\frac{\Omega^2}{16 E_r}$. In \eqref{eq_intham_nonsymmetric}, we have excluded the terms $\propto g_2\Omega^2$, since typically $\abs{g_2} \ll g_0$. Notice that, even in the case of $SU(3)$-symmetric interactions (i.e. $g_2=0$), $\ham_\mathrm{A}$ includes SOC-induced spin-changing collision processes with a spin-mixing rate $\tilde{g}_2$. 

\textit{Three-mode model.---} We now restrict ourselves to the case in which $\ham_\mathrm{A}$ can be treated as a perturbation over the symmetric part $\ham_\mathrm{S}$. We assume that the dynamics is then well described by a three-mode model. It includes three eigenmodes of $\ham_\mathrm{S}$, labeled as $\ket{\phi_{-1}}$, $\ket{\phi_{0}}$ and $\ket{\phi_{1}}$, which have a quasi-momentum distribution centered at the vicinity of $\vec{k}_{-1}$, $\vec{k}_{0}$ and $\vec{k}_{1}$, respectively. By introducing the associated bosonic operators $\hat{b}_{-1}$, $\hat{b}_0$ and $\hat{b}_{1}$, we truncate the field operators to $\hat{\varphi}_{i}\dgr(\vec{r}) \sim \phi_{i}^{*}(\vec{r}) \hat{b}\dgr_{i}$. We call the three modes, $\ket{\phi_j}$, pseudospin states. Finally, dropping the terms that only depend on the total number of particles, $N$, we obtain the following one-axis-twisting Hamiltonian
\begin{equation}\label{eq_eff_spin_ham}
\ham_\mathrm{eff} = \frac{\lambda}{2N}\hat{L}^2 - \frac{\lambda-{g}_2n}{2N}\hat{L}_z^2 + \delta \hat{L}_z +\tilde{\epsilon}\hat{L}_{zz},
\end{equation}
where we introduce the collective pseudospin operators $\hat{L}_{x,y,z} = \sum_{\mu\nu}\hat{b}_{\mu}\dgr (\hat{F}_{x,y,z})_{\mu\nu} \hat{b}_{\nu}$ and $\hat{L}_{zz} =\sum_{\mu\nu}\hat{b}_{\mu}\dgr (\hat{F}_{z}^2)_{\mu\nu}\hat{b}_{\nu}$. Here, $\lambda = (\tilde{g}_2 + g_2)n$, where $n$ is the mean density of the gas \footnote{Since the spinor modes $\ket{\phi_j}$ are determined through the symmetric Hamiltonian \eqref{eq_ham_symmetric}, we have that $|\phi_i(\vec{r})| = |\phi_j(\vec{r})|$ for all $i,j=-1,0,1$. Thus, within the subspace spanned by these three modes, the mean density of the gas is simply given by $n = N \int d\vec{r} |\phi_0(\vec{r})|^4$.}.

Since $[\ham_\mathrm{eff}, \hat{L}_z ] = 0$, the total magnetization is preserved by $\ham_\mathrm{eff}$. Within the zero magnetization subspace  (where $\hat L_z = 0$), the effective Hamiltonian \eqref{eq_eff_spin_ham} reduces to 
\begin{equation}\label{eq_spin_ham_mz0}
\ham_0 = \lambda\frac{\hat{L}^2}{2N} + \tilde{\epsilon}\hat{L}_{zz}.
\end{equation}
Hamiltonian \eqref{eq_spin_ham_mz0} describes the nonlinear coherent spin dynamics in a spin-1 BEC, in which the density-dependent spin-symmetric interaction dominates \cite{Law-1998}. In the SOC-based realization of \eqref{eq_spin_ham_mz0} we propose here, we can control the spin-mixing parameter $\lambda$ independently of the density of the gas by adjusting $\Omega$. That is, SOC BECs provide a novel platform for designing entanglement protocols and studying dynamical phase transitions.

\textit{Dynamical preparation of stripe states.---} The phase diagram of Hamiltonian \eqref{eq_spin_ham_mz0} in the $\Omega-\epsilon$ plane is shown in Fig.\,\ref{Fig_eff_low_en_phys}(c1), where we use the expressions for $\lambda(\Omega)$ and $\tilde{\epsilon}(\Omega,\epsilon)$. We consider $^{87}$Rb, with $g_2/g_0 = -0.0047$  \cite{Stamper-Kurn-2013}, and density $n=7.5\cdot10^{13}$ cm$^{-3}$. We now use this effective description to design a protocol to prepare dynamically the stripe phase of the dressed gas, which we later test numerically. For $\Omega > \Omega_c = 4 E_r\sqrt{\abs{g_2}/g_0}$, the diagram is equivalent to that of an antiferromagnetic spinor gas without SOC, $\lambda >0$. The ground state is then either in a polar (P) phase, where all the atoms occupy the $\ket{\phi_0}$ state, or in a twin-Fock (TF) phase, in which the ground state approximates the spin-$1/2$ balanced Dicke state $\frac{1}{(N/2)!}(\hat{b}_{-1}\dgr)^{N/2}(\hat{b}_{1}\dgr)^{N/2} \ket{0}$. The phase transition between the two phases is found along $\tilde{\epsilon}(\Omega) = 0$. At $\Omega = \Omega_c$, the effective and the intrinsic spin-mixing dynamics mutually compensate, with $\tilde{g}_2 = -g_2$, yielding $\lambda=0$. For $\Omega < \Omega_c$, the effective spin dynamics is ferromagnetic, $\lambda < 0$. Then, dressed spin interactions tend to maximize the total spin, resulting in ground state with a non-vanishing transverse magnetization. This spontaneous breaking of the SO(2) symmetry of the system \cite{Sadler-nature-2006} gives rise to the so-called broken-axisymmetry (BA) phase \cite{Murata-2007} in between the P and TF phases. The two transitions take place at $\tilde{\epsilon} = \pm 2\lambda$ in the thermodynamic limit. The three phases meet at the tricritical point $C_{F}$, at $\Omega = 4 E_r\sqrt{\abs{g_2}/g_0}$ and $\epsilon = g_2/g_0$. 

Remarkably, the BA phase of the effective model corresponds to the super-solid like ferromagnetic stripe (FS) phase of the spin-1 SOC gas diagram, described in detail in \cite{Martone-2016}. The FS phase is characterized by the presence of spatial density modulations that are proportional to $\Omega$. When $\abs{g_2}$ is small, as in $^{87}$Rb, such phase is only favored in a very narrow region in parameter space, which makes its experimental realization challenging. Alternatively, the ferromagnetic landscape can be probed in the most-excited manifold of $\ham_0$ in the antiferromagnetic regime, given that $\ham_0 (\lambda, \tilde{\epsilon}) = -\ham_0 (-\lambda, -\tilde{\epsilon})$.  In  Fig.\,\ref{Fig_eff_low_en_phys}(c2), we show the phase diagram for the most excited state of $\ham_0$. It displays the same phases as the ground state, but with the phase boundaries redefined. In the excited-state diagram, the predicted BA phase occurs for a much broader range of parameters. Notably, at the P-BA and BA-TF transitions, the energy gap between the two most excited states scales weakly with the total number of particles as $\propto \lambda N^{-1/3}$. This facilitates the quasi-adiabatic driving through both phase transitions in workable time scales even when the number of particles is large. This feature was exploited in \cite{Luo-2017} and \cite{Zou-2018} to generate macroscopic TF and BA states, respectively, in small $^{87}\text{Rb}$ spinor condensates. 

Following the dressed-spinor description, we propose to prepare the FS phase in the most excited phase diagram of the effective model by driving an initially polarized state across the P-BA quantum phase transition therein. The loading can be easily achieved from an undressed condensate in the $m_f=0$ spin state by adiabatically turning up $\Omega$, while setting $\tilde{\epsilon} < -2\lambda$. The excited phase diagram can then be probed by varying $\epsilon$ and $\Omega$. Since here the stripe phase occurs at larger $\Omega$, it exhibits a larger contrast of the density modulations, when compared to its ground-state counterpart. 

\begin{figure}[t!]
\includegraphics[width=0.95\linewidth]{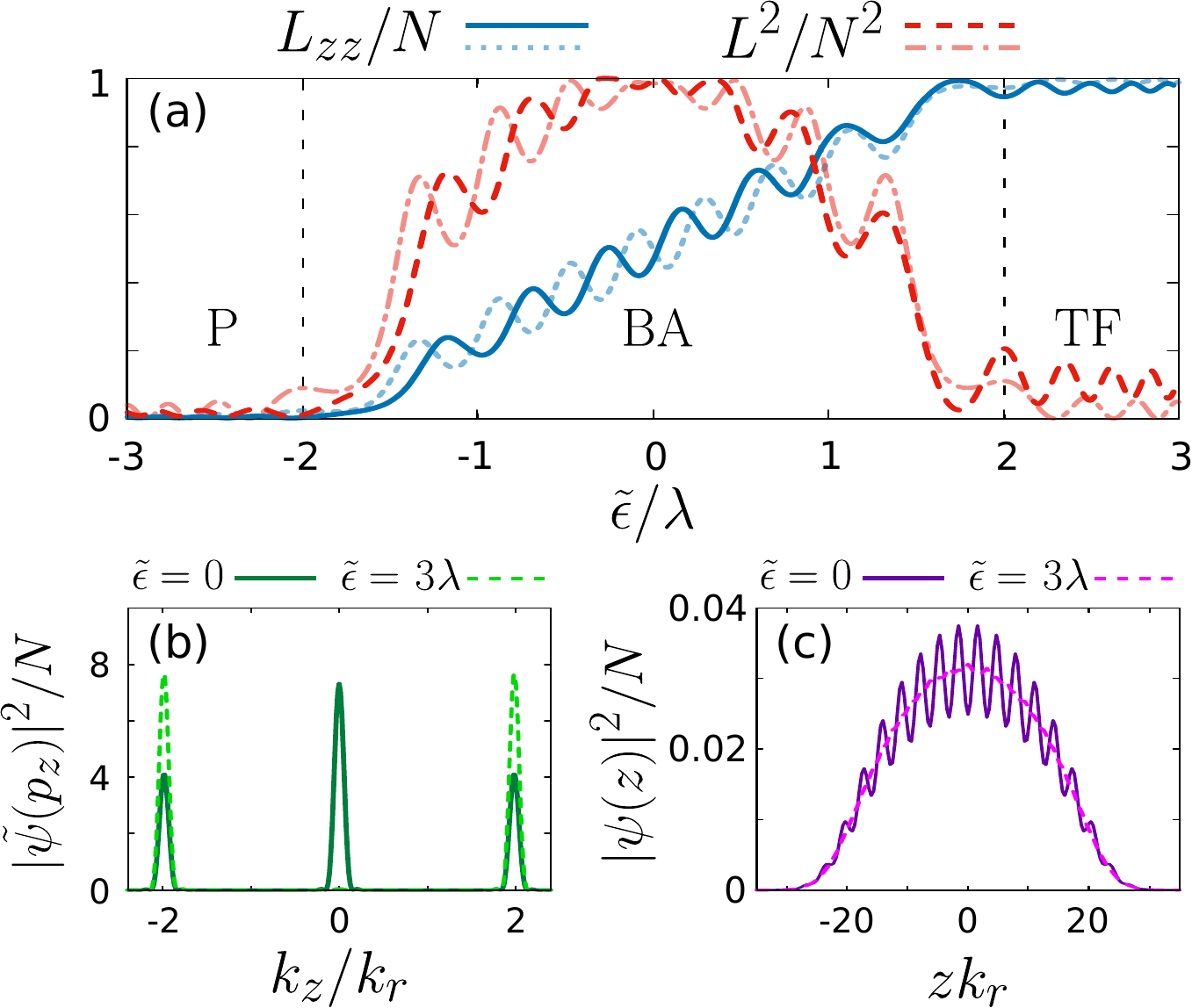}
\caption{(Color online) \textbf{Crossing quantum phase transitions in an excited state.} (a) $L_{zz}$ (solid blue) and $L^2$ (dashed red) as a function of $\tilde{\epsilon}$ for a state initially prepared at $b_{\pm1} = \sqrt{50}$ and $b_0 = \sqrt{N-100}$, with $N(0)=10^4$ and $\hbar\omega_\mathrm{t}= 2\pi\cdot 140\,$Hz.The state is evolved under the GPE while driving $\tilde{\epsilon}$ from $-3\lambda$ to $3\lambda$, keeping $\Omega = 0.65 E_r$, following the red dashed path in Fig.\,\ref{Fig_eff_low_en_phys}(c2). The total drive time is set to $\tau_d = 8h/\lambda$. The corresponding results obtained with simulations of the three-mode model \eqref{eq_spin_ham_mz0} are shown in light colors. (b) Quasi-momentum density $|\tilde{\vec{\psi}} (p_z)|^2$ of the driven state at $\tilde{\epsilon}=0$ (solid dark green) and $\tilde{\epsilon}= 3 \lambda$ (dashed light green). (c) Corresponding density profiles at $\tilde{\epsilon}=0$ (solid purple) and $\tilde{\epsilon}=3 \lambda$ (dashed pink).}\label{Fig_mean_field_results} 
\end{figure}

To derive our protocol, we assume the validity of the three-mode truncation that leads to Hamiltonian \eqref{eq_spin_ham_mz0}. To assess the extent of such truncation, which is equivalent to the single-spatial mode approximation in spinor condensates \cite{You-2002-PRA-SMA}, we simulate the protocol with the Gross–Pitaevskii equation (GPE) for the full dressed gas, $i\hbar \dot{\psi}_j = \delta \mathcal E/\delta \psi_j^*$, with $\mathcal E \!=\! \vec{\psi}^{*}\!\! \left(\hat{\mathcal{H}}_{\mathrm{k}} \!+\! V_\mathrm{t}\right) \vec{\psi} + \frac{g_0}{2}\abs{\vec{\psi}}^4 \!+\! \frac{g_2}{2}\sum_j (\vec{\psi}^{*} \hat{F}_j \vec{\psi})^2 $, using the XMDS2 library \cite{Dennis-CPC-2013} (See Supplemental Material for more details). We label the three self-consistent modes around $\vec{k}_j$ as $\vec{\phi}_{j}$, which are calculated via imaginary time evolution of the GPE, and define $b_j = \int d\vec{r} \vec{\phi}^*_{j}(\vec{r})\cdot\vec{\psi}(\vec{r})$. As a reference, we consider similar conditions to those described in \cite{Zou-2018}, with small $^{87}$Rb condensates in the $F=1$ hyperfine manifold at $n \sim 7.5\cdot 10^{13}$\,cm$^{-3}$, and take $E_r/\hbar = 2\pi \cdot 3680\,$Hz, $k_r = 7.95\cdot 10^{6}$\,m$^{-1}$ and $g_0 k_r^3 = 1.066\,E_r$. Note that in the proposed protocol, the state is initially prepared in the Fock state $\frac{1}{\sqrt{N!}}(\hat{b}_0^\dagger)^N\ket{0}$. In these conditions, the dynamics is dominated by quantum fluctuations \cite{Klempt-prl-2010, evrard-arxiv-2021-coherent}, and the mean field description is expected to be inaccurate. Instead, we set the initial state to a coherent state with $0  <b_{\pm 1} \ll N $. 

In Fig.\,\ref{Fig_mean_field_results} we show the results for a drive along the red dashed path drawn in the excited state diagram from Fig.\,\ref{Fig_eff_low_en_phys}(c2). The drive is obtained with $\delta=0$, $\omega_\mathrm{t}= 2\pi\cdot 140\,$Hz and $N=10^4$. We set $\Omega=0.65 E_r$, and the initial state to $b_{\pm1} = \sqrt{50}$ and $b_0 = \sqrt{N-100}$. In Fig.\,\ref{Fig_mean_field_results}(a) we plot the collective pseudospin $L^2 = \sum_j (\sum_{\mu\nu}b_{\mu}^* (\hat{F}_{j})_{\mu\nu} b_{\nu})^2$ and the tensor magnetization $L_{zz} =\sum_{\mu\nu}b_{\mu}^* (\hat{F}_{z}^2)_{\mu\nu}b_{\nu}$ as a function of $\tilde{\epsilon}/\lambda$. The state is time evolved following the linear ramp $\tilde{\epsilon}(t) = 3\lambda(2 t/\tau_d-1)$, with $\tau_d = 8 h/\lambda$, that crosses both transitions at $\tilde{\epsilon} \sim \pm 2\lambda$. In the BA phase, the tensor magnetization $\hat{L}_{zz}$ increases homogeneously with $\tilde{\epsilon}/\abs{\lambda}$, and the total spin $\hat{L}$ peaks at $\tilde{\epsilon}=0$, in agreement with the effective model (see Fig.\ref{Fig_eff_low_en_phys}(c)). For comparison, the results obtained from the direct simulation of the three-mode Hamiltonian \eqref{eq_spin_ham_mz0} are shown in light colors. In Fig.\,\ref{Fig_mean_field_results}(b) we plot the momentum-space density at the middle and at the end of the drive, in which the state approaches a BA state and a TF state, respectively. The corresponding density profiles are shown in Fig.\,\ref{Fig_mean_field_results}(c). As expected, the excited BA phase exhibits large density modulations along the direction of the Raman beams.

\textit{Experimental considerations.---} Finally, we assess the robustness of the preparation by incorporating atom loss and heating mechanisms into the simulations of the GPE. We naively model the noise in $\delta$ and $\epsilon$ with sinusoidal signals of frequency $50\,$Hz and amplitude $300$\,Hz and $2.5$\,Hz, respectively. We consider $\Omega$ to be stable during the drive, but to have a calibration uncertainty of $125$\,Hz in each realization. These amplitudes are compatible with a magnetic bias field instability of $\sim 0.5$ mG and a relative uncertainty of $\pm 5\%$ in $\Omega$, within the stabilities reached in experiments with $^{87}$Rb \cite{Lin-2011, Campbell-2016, Xu-2019}. At the same time we consider a $10\%$ uncertainty in the number of atoms initially in the condensate, and the population to decay as $N(t) = N(0)\exp(-\gamma t)$, with $\gamma = 3.33\,$s$^{-1}$, which is compatible with the lifetime of spin-1 Raman-dressed BECs for $\Omega < E_r$ \cite{Campbell-2016, Anderson-PRR-2019}.

\begin{figure}[t]
\includegraphics[width=0.99\linewidth]{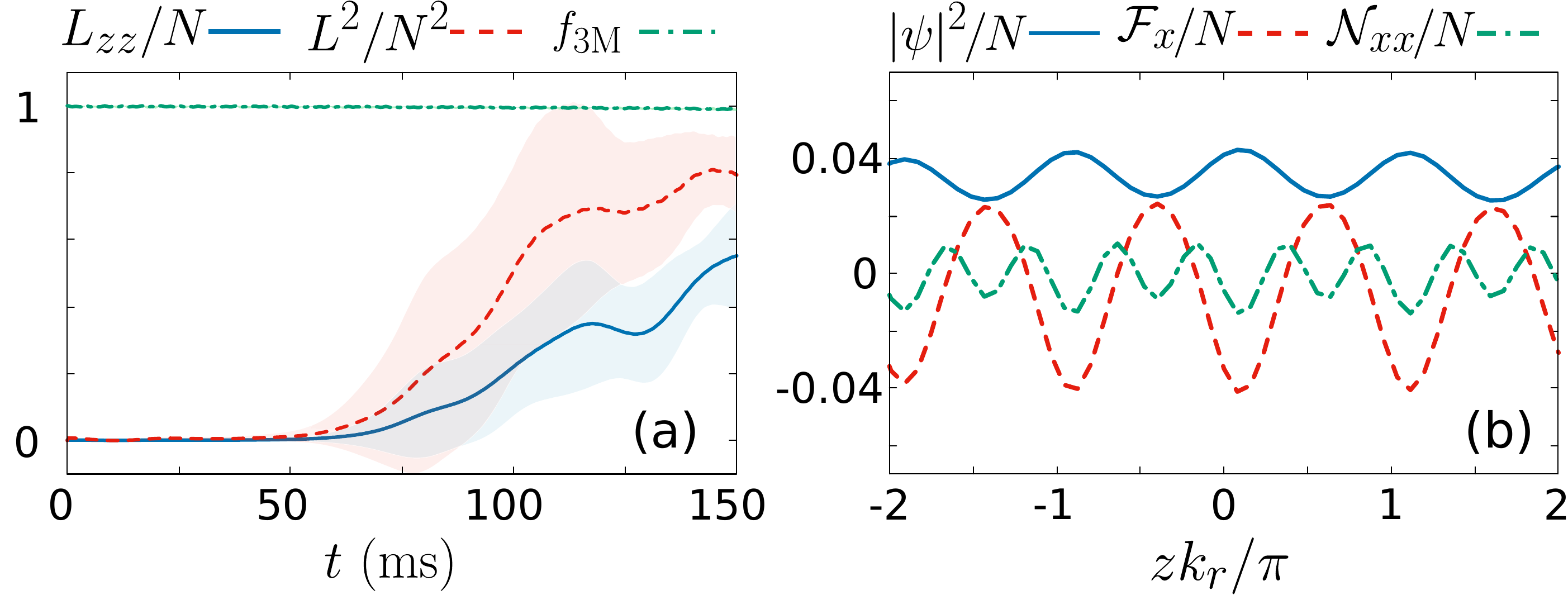}
\caption{(Color online) \textbf{Robust preparation of FS states.} 
(a) $L_{zz}$ (solid blue), $L^2$ (dash-dotted red) and $f_\mathrm{3M}$ as a function of time for a state initially prepared at $b_{\pm1} = \sqrt{10}$ and $b_0 = \sqrt{N-20}$, with $N(0)=10^4$ and $\hbar\omega_\mathrm{t}= 2\pi\cdot 140\,$Hz. The state is evolved under the GPE while driving $\tilde{\epsilon}$ from $-3\lambda$ to $0$ by linearly increasing $\Omega$ from $0.65 E_r$ to $0.767 E_r$, following the blue dashed-dotted path in Fig.\,\ref{Fig_eff_low_en_phys}(c2). The parameters of the GPE are subject to random fluctuations that simulate experimental noise, as described in the main text, and the values depicted are averaged over 20 realizations. The shadowed regions indicate the associated standard deviations. (b) Longitudinal density $|\vec{\psi}|^2$ (solid blue), spin density $\mathcal{F}_x$ (dashed red) and nematic density $\mathcal{N}_{xx}$ (dashed-dotted green) at $t=150$ ms from a single realization of the drive. }\label{Fig_FS_preparation} \end{figure}

In these conditions, we simulate a drive following the blue dashed-dotted path drawn in the excited state diagram from Fig.\,\ref{Fig_eff_low_en_phys}(c2). Along the path, $\epsilon$ is kept fixed while $\Omega$ is linearly ramped up. In this way, $\lambda$ is increased as $\tilde{\epsilon}$ approaches to $0$. Such tunability of the SOC-mediated spin-mixing allows to reduce the preparation time while retaining a high robustness. At the same time, at larger $\Omega$, the contrast of the stripes is further enhanced. In Fig.\,\ref{Fig_FS_preparation}(a), we plot $L_{zz}$, $L^2$ and the fraction of atoms that remain within the three-mode subspace, $f_\mathrm{3M} = \frac{1}{N^2}\sum_j \abs{b_j}^2$, averaged over 20 of drives. The P-BA transition is well captured, with $f_\mathrm{3M} \sim 0.99$ by the end of the drives. Finally, in Fig.\,\ref{Fig_FS_preparation}(b) we plot the longitudinal density $\abs{\vec{\psi}}^2$, the spin density $\mathcal{F}_x= \vec{\psi}^*\hat{F}_{x}\vec{\psi}$ and the nematic density $\mathcal{N}_{xx} = \vec{\psi}^*(2/3-\hat{F}_{x}^2)\vec{\psi} $  at $\tilde\epsilon = 0$ for a single realization of the drive. As predicted by the effective model, the prepared state exhibits the characteristic properties of FS states, with large spatial modulations along the direction of the Raman beams. The FS phase can be distinguished from antiferromagnetic stripe phases from the periodicity of the modulations, with the particle density and the spin densities having periodicity $2\pi/\abs{\vec{k_1}} $, and the nematic densities containing harmonic components both with period $2\pi/\abs{\vec{k_1}} $ and $\pi/\abs{\vec{k_1}}$. As a final remark, we note that the preparation could be optimized further by employing reinforcement learning techniques, as recently demonstrated in \cite{Guo-arXiv-2020}.

\textit{Conclusions.---} In summary, we have shown that, for weak Raman coupling and interactions, a Raman-dressed spin-1 BEC is equivalent to an artificial spinor BEC with tunable nonsymmetric spin interactions. A ferromagnetic gas like $^{87}$Rb can be turned to antiferromagnetic by light dressing, and the stability of the FS phase understood in these terms. We have used such insight to propose the preparation of FS phases by driving an initially polarized state through a quantum phase transition in an excited state of the Raman-dressed gas. In the excited-state phase diagram, the FS phase is broader, and both the energy gap and the density modulation contrast are larger. These features enable a robust preparation of the state and ease the detection of its supersolid properties, e.g. by probing its spectrum of excitations \cite{Li-2013, Geier-arXiv-2021}.

Our dressed-base description of Raman-coupled spinor gases suggests new directions for probing nonequilibrium phenomena, as in \cite{Prufer-nature-2018, Prufer-natphys-2020}, with light-dressed spinor gases of alkali and non-alkali \cite{Chalopin-natphys-2020} atoms. Remarkably, the FS phase corresponds to the BA entangled phase of the artificial spinor gas: its preparation may thus lead to the generation of macroscopic entanglement in momentum space, {\it cf.} \cite{anders2020momentum}. Likewise, the map introduces SOC gases as a novel platform to study dynamical and excited-state quantum phase transitions. The FS phase of the spin-1 gas can be understood as an excited-state quantum phase through its connection with undressed collisional spin dynamics \cite{Feldmann-arxiv-2020-ESQPT}. This precise connection will be explored in an upcoming work \cite{Cabedo-ESQPTs-2021-workinprogress}.

\begin{acknowledgments}
We thank J. Mompart and V. Ahufinger for useful discussions and L. Tarruell for insightful discussions on experimental aspects of the Raman coupled BEC. A.C. thanks G. Juzeliunas for discussions on Raman coupled spinor BEC during his stay at Institute of Theoretical Physics and Astronomy of University of Vilnius, supported by the COST action 16221, The Quantum Technologies with Ultracold atoms. J. Cabedo and A.C. acknowledge support from the Ministerio de Economía y Competividad MINECO (Contract No. FIS2017-86530-P), from the European Union Regional Development Fund within the ERDF Operational Program of Catalunya (project QUASICAT/QuantumCat), and from Generalitat de Catalunya (Contract No. SGR2017-1646). A.C. acknowledges support from the UAB Talent Research program. J. Claramunt acknowledges partial support from the research funding Brazilian agency CAPES and the European Research Council (ERC) under the European Union’s Horizon 2020 research and innovation programme (Grant agreement No. 805495).
\end{acknowledgments}



%

\pagebreak
\widetext
\begin{center}
\textbf{\large Supplemental Material}
\end{center}
\setcounter{equation}{0}
\setcounter{figure}{0}
\setcounter{table}{0}
\makeatletter
\renewcommand{\theequation}{S\arabic{equation}}
\renewcommand{\thefigure}{S\arabic{figure}}

In this supplementary document we include the detailed derivation of the low-energy Hamiltonian introduced in the main text. We also provide additional insights on the approach taken to assess the validity of the three-state model derived, and on its robustness.

\section{Effective low-energy theory}

Here we detail the derivation of the effective low-energy theory presented in the main text for weakly-coupled Raman-dressed spin-1 BECs, with Rabi frequency $\Omega< 1$ (in units of recoil energy). We restrict ourselves to a regime in which the linear and quadratic Zeeman terms, denoted by $\delta$ and $\epsilon$ respectively, are also small, and set $\abs{\delta},\abs{\epsilon}\ll 1$. In this regime, the low-energy landscape only involves the dressed states located around the three minima of the dispersion band. Thus, we set a cut-off $\Lambda \ll 1$ (in units of $k_r$) to the momentum spread $p$ around each minimum, so that $|p| < \Lambda$. Under these conditions, we use second order perturbation theory to express the bare fields $\hat \psi_i$ in terms of the lowest-band dressed-state fields $\hat \varphi_j$ around the center band minimum
\begin{align}
\hat \psi_{0}( p) 
&= \left(1 - \frac{\Omega^2}{64} \left(1 - \frac{\epsilon}2 + O((\Lambda+ \frac{\epsilon+\delta}4)^2)\right)\right)\,  \hat \varphi_{0} (p) + O\left( (\frac{\Omega}{8(1-\Lambda)})^3\right), \cr
\hat \psi_{\pm 1}(p) 
&= -\frac{\Omega}8 \left(1 -\frac{\epsilon \pm \delta \mp 4p}4 + O((\Lambda+ \frac{\epsilon+\delta}4)^2)\right)\, \hat\varphi_{0}(p) 
+ O\left( (\frac{\Omega}{8(1-\Lambda)})^3\right),
\cr
\end{align}
and in right/left band minima
\begin{align}\label{dressed_basis_edge}
\hat \psi_{\pm 1}(\pm 2 + p) 
&= \left( 1- \frac 12 \left(\frac{\Omega}8\right)^2 \left(1 +\frac{\epsilon \pm \delta \mp 4p}2+ O((\Lambda+ \frac{\epsilon+\delta}4)^2)\right)\right)\,\hat\varphi_{\pm 1}(p)
+ O\left( (\frac{\Omega}{8(1-\Lambda)})^3\right), \cr
\hat \psi_0(\pm 2+p) 
&= -\frac{\Omega}8 \left(1 +\frac{\epsilon \pm \delta \mp 4p}4+ O((\Lambda+ \frac{\epsilon+\delta}4)^2)\right)\,\hat\varphi_{\pm 1}(p)
+ O\left( (\frac{\Omega}{8(1-\Lambda)})^3\right),\cr
\hat \psi_{\mp 1}(\pm 2 + p) &= \frac{\Omega^2/16}{\left((16 + p^2 \pm \delta \pm 8p\right) \left(1-\frac{\epsilon \mp \delta \pm 4p}{4} \right)} \, \hat\varphi_{\pm 1}(p) + O\left( (\frac{\Omega}{8(1-\Lambda)})^3\right),
\cr
\end{align}
respectively. We made explicit only the dependence on momentum along the direction of the recoil momentum transfer. Notice that the positions of the edge band minima are actually shifted from $\pm 2$ by a small amount proportional to $\Omega^2$. Still, up to second order in $\Omega$, these shifts do not contribute to expressions \eqref{dressed_basis_edge}, and hence are not included. Note that the last term of the above expressions can be neglected since it contributes to the interactions at fourth order in $\frac{\Omega}{8(1-\Lambda)}$. As shown below, due to momentum conservation, the nontrivial contributions to the interacting Hamiltonian involve only the first order terms in the above expressions, while the second order just renormalize the symmetric interactions.

We adopt the notation short cuts
\begin{equation}\label{eq-shortcut1}
\int\int \hat\psi_ a^\dagger\hat\psi_b^\dagger \hat\psi_a \hat\psi_b\equiv \frac g2 \int {\rm d} r \int\prod_{j=1}^4 \frac{{\rm d}^3 k_j}{(2\pi)^3} e^{i {\bf r}\cdot ({\bf k}_1+{\bf k}_2-{\bf k}_3-{\bf k}_4)}
 \hat\psi_a^\dagger({\bf k}_1)\hat\psi_b^\dagger({\bf k}_2) \hat\psi_a({\bf k}_3)\hat\psi_b({\bf k}_4),
\end{equation}
and
\begin{equation}\label{eq-shortcut2}
\int\int \hat\varphi_a^\dagger\hat\varphi_b^\dagger \hat\varphi_a \hat\varphi_b\equiv \frac g2 \int {\rm d} r \int_{-\Lambda}^{\Lambda} \prod_{j=1}^4 \frac{{\rm d}^3 p_j}{(2\pi)^3} e^{i {\bf r}\cdot ({\bf p}_1+{\bf p}_2-{\bf p}_3-{\bf p}_4)}
 \hat\varphi_a^\dagger({\bf p}_1)\hat\varphi_b^\dagger({\bf p}_2) \hat\varphi_a({\bf p}_3)\hat\varphi_b({\bf p}_4).
\end{equation} 

When the interaction operators are evaluated on the low-energy states, it follows that
\begin{align}\label{int-symmetric1}
\int\int \hat\psi_{\pm}^\dagger\hat\psi_{\pm}^\dagger \hat\psi_{\pm} \hat\psi_{\pm}=& \int\int \left(1-\frac{\Omega^2}{32} \left(1 +\frac{\epsilon\pm\delta}2 \mp \frac{p_1+p_2+p_3+p_4}2 + O((\Lambda+ \frac{\epsilon+\delta}4)^2)\right)\right)\,
\hat\varphi_{\pm}^\dagger\hat\varphi_{\pm}^\dagger \hat\varphi_{\pm} \hat\varphi_{\pm} \cr
&+\frac{\Omega^2}{16}\int\int \left(1 -\frac{\epsilon\pm\delta}2 \pm (p_2+p_4) +  O((\Lambda+ \frac{\epsilon+\delta}4)^2)\right)\,
\hat\varphi_{\pm}^\dagger\hat\varphi_{0}^\dagger \hat\varphi_{\pm} \hat\varphi_{0},\cr
\end{align}
\begin{align}\label{int-symmetric2}
\int\int \hat\psi_{0}^\dagger\hat\psi_{0}^\dagger \hat\psi_{0} \hat\psi_{0}=& \int\int \left(1 - \frac{\Omega^2}{16} \left(1- \frac{\epsilon}2 + O((\Lambda+ \frac{\epsilon+\delta}4)^2)\right)\right)\,
\hat\varphi_{0}^\dagger\hat\varphi_{0}^\dagger \hat\varphi_{0} \hat\varphi_{0} \cr
&+\frac{\Omega^2}{16}\int\int \left(1 +\frac{\epsilon + \delta}2 - (p_1+p_3) + O((\Lambda+ \frac{\epsilon+\delta}4)^2)\right)\,\hat\varphi_{+}^\dagger\hat\varphi_{0}^\dagger \hat\varphi_{+} \hat\varphi_{0}\cr
&+\frac{\Omega^2}{16}\int\int \left(1 +\frac{\epsilon - \delta}2 + (p_1+p_3) + O((\Lambda+ \frac{\epsilon+\delta}4)^2)\right)\,\hat\varphi_{-}^\dagger\hat\varphi_{0}^\dagger \hat\varphi_{-} \hat\varphi_{0}\cr
&+ \frac{\Omega^2}{32}\int\int \left(1 +\frac{\epsilon}2 - (p_1-p_2)+ O((\Lambda+ \frac{\epsilon+\delta}4)^2)\right)\,\hat\varphi_{+}^\dagger\hat\varphi_{-}^\dagger\hat\varphi_{0} \hat\varphi_{0}\cr
&+ \frac{\Omega^2}{32}\int\int \left(1 +\frac{\epsilon}2 - (p_3-p_4)+ O((\Lambda+ \frac{\epsilon+\delta}4)^2)\right)\,\hat\varphi_{0}^\dagger\hat\varphi_{0}^\dagger\hat\varphi_{+} \hat\varphi_{-},\cr
\end{align}
\begin{align}\label{int-symmetric3}
\int\int \hat\psi_{\pm}^\dagger\hat\psi_{0}^\dagger \hat\psi_{\pm} \hat\psi_{0}=& \int\int \left(1 - \frac{\Omega^2}{64} \left(3 - \frac{\epsilon \mp \delta}2 \mp (p_1+p_3)
 + O((\Lambda+ \frac{\epsilon+\delta}4)^2)\right)\right)\,
\hat\varphi_{\pm}^\dagger\hat\varphi_{0}^\dagger \hat\varphi_{\pm} \hat\varphi_{0}\cr
&+\frac{\Omega^2}{64}\int\int \left(1 -\frac{\epsilon\pm\delta}2 \pm (p_1+p_3) +  O((\Lambda+ \frac{\epsilon+\delta}4)^2)\right)\,
\hat\varphi_{0}^\dagger\hat\varphi_{0}^\dagger \hat\varphi_{0} \hat\varphi_{0}\cr
&+\frac{\Omega^2}{64}\int\int \left(1 \pm (p_1-p_2) +  O((\Lambda+ \frac{\epsilon+\delta}4)^2)\right)\,
\hat\varphi_{0}^\dagger\hat\varphi_{\pm}^\dagger \hat\varphi_{\pm} \hat\varphi_{0}\cr
&+\frac{\Omega^2}{64}\int\int \left(1 \pm (p_3-p_4) +  O((\Lambda+ \frac{\epsilon+\delta}4)^2)\right)\,
\hat\varphi_{0}^\dagger\hat\varphi_{\pm}^\dagger \hat\varphi_{0} \hat\varphi_{\pm}\cr
&+\frac{\Omega^2}{64}\int\int \left(1 +\frac{\epsilon \pm \delta}2 \mp (p_2+p_4) + O((\Lambda+ \frac{\epsilon+\delta}4)^2)\right)\,\hat\varphi_{\pm}^\dagger\hat\varphi_{\pm}^\dagger \hat\varphi_{\pm} \hat\varphi_{\pm}\cr
&+\frac{\Omega^2}{64}\int\int \left(1 +\frac{\epsilon \mp + \delta}2 \pm (p_2+p_4) + O((\Lambda+ \frac{\epsilon+\delta}4)^2)\right)\,\hat\varphi_{\pm}^\dagger\hat\varphi_{\mp}^\dagger \hat\varphi_{\pm} \hat\varphi_{\mp}\cr
&+\frac{\Omega^2}{64} \left(1 \mp \delta/2 \pm (p_2+p_3) + O((\Lambda+ \frac{\epsilon+\delta}4)^2)\right)\,\hat\varphi_{\pm}^\dagger\hat\varphi_{\mp}^\dagger \hat\varphi_{0} \hat\varphi_{0} \cr
&+\frac{\Omega^2}{64} \left(1 \mp \delta/2 \pm (p_1+p_4) + O((\Lambda+ \frac{\epsilon+\delta}4)^2)\right)\,\hat\varphi_{0}^\dagger\hat\varphi_{0}^\dagger \hat\varphi_{\pm} \hat\varphi_{\mp}, 
\end{align}
\begin{align}\label{int-symmetric4}
\int\int \hat\psi_{\pm}^\dagger\hat\psi_{\mp}^\dagger \hat\psi_{\pm} \hat\psi_{\mp}=
& \int\int \left(1-\frac{\Omega^2}{32} \left(1+\frac{\epsilon}2 \mp\frac{p_1-p_2+p_3-p_4}2 + O((\Lambda+ \frac{\epsilon+\delta}4)^2)\right)\right)\, 
\hat\varphi_{\pm}^\dagger\hat\varphi_{\mp}^\dagger \hat\varphi_{\pm} \hat\varphi_{\mp}\cr
&+\frac{\Omega^2}{64}\int\int \left(1 -\frac{\epsilon\pm\delta}2 \pm (p_1+p_3) +  O((\Lambda+ \frac{\epsilon+\delta}4)^2)\right)\,
\hat\varphi_{0}^\dagger\hat\varphi_{\mp}^\dagger \hat\varphi_{0} \hat\varphi_{\mp}\cr
&+\frac{\Omega^2}{64}\int\int \left(1 -\frac{\epsilon\mp\delta}2 \mp (p_2+p_4) +  O((\Lambda+ \frac{\epsilon+\delta}4)^2)\right)\,
\hat\varphi_{\pm}^\dagger\hat\varphi_{0}^\dagger \hat\varphi_{\pm} \hat\varphi_{0}\cr
&+\frac{\Omega^2}{64}\int\int \left(1 -\frac{\epsilon}2 \pm (p_1-p_2) +  O((\Lambda+ \frac{\epsilon+\delta}4)^2)\right)\,
\hat\varphi_{0}^\dagger\hat\varphi_{0}^\dagger \hat\varphi_{\pm} \hat\varphi_{\mp} \cr
&+\frac{\Omega^2}{64}\int\int \left(1 -\frac{\epsilon}2 \pm (p_3-p_4) +  O((\Lambda+ \frac{\epsilon+\delta}4)^2)\right)\,
\hat\varphi_{\pm}^\dagger\hat\varphi_{\mp}^\dagger \hat\varphi_{0} \hat\varphi_{0}.
\end{align}
Inserting \eqref{int-symmetric1}-\eqref{int-symmetric4} into the symmetric contribution to the interacting Hamiltonian $\hat V_s$, we get
\small
\begin{align}\label{symm_interaction_dressed}
\hat V_s &= \int\int \left(\sum_{a=-1,0,+1} \left(\hat \psi_a^\dagger\hat\psi_a^\dagger \hat\psi_a \hat\psi_a +2\sum_{b>a} \hat\psi_a^\dagger\hat\psi_b^\dagger \hat\psi_a \hat\psi_b\right)\right)  \cr
&=  \int\int \left(\sum_{a=-1,0,+1} \left(\hat \varphi_a^\dagger\hat\varphi_a^\dagger \hat\varphi_a \hat\varphi_a +2\sum_{b>a} \hat\varphi_a^\dagger\hat\varphi_b^\dagger \hat\varphi_a \hat\varphi_b\right)\right) \cr
&+ \frac{\Omega^2}8 \int\int \left( \left(\hat\varphi_{+1}^\dagger \hat\varphi_{+1} + \hat\varphi_{-1}^\dagger \hat\varphi_{-1}\right)\hat\varphi_{0}^\dagger \hat\varphi_{0} + \left(\hat \varphi_{+1}^\dagger \hat\varphi_{-1}^\dagger \hat\varphi_{0} \hat\varphi_{0} + \textit{H.c.} \right)  + O((\Lambda + \frac{\epsilon + \delta}{4})^2)\right) \cr
& + \frac{\Omega^2}{16} \int\int \left((p_2-p_1+p_4-p_3)\left(\hat\varphi_{+1}^\dagger \hat\varphi_{0}^\dagger \hat\varphi_{+1} \hat\varphi_{0} -\hat\varphi_{-1}^\dagger \hat\varphi_{0}^\dagger \hat\varphi_{-1} \hat\varphi_{0} + (\hat\varphi_{+1}^\dagger \hat\varphi_{-1}^\dagger \hat\varphi_{0} \hat\varphi_{0} + \textit{H.c.} )\right) + O((\Lambda + \frac{\epsilon + \delta}{4})^2)\right).
\end{align}
\normalsize
The last term in \eqref{symm_interaction_dressed} contains a correction to the spin-mixing contribution that depends linearly on the momentum. However, its value is bounded by the cutoff in the momentum spread around the wells. Since $|p_i-p_j| < 2 \Lambda \ll 1$, for simplicity we neglect such correction to the interacting Hamiltonian.

Finally, considering that, for $\abs{p}>\Lambda$, the fields $\hat{\varphi}_j(p)$ vanish when acting on the low energy subspace, we can formally remove the cut-off in the integration and perform the Fourier transform. By doing so, we obtain the expression introduced in the main text for the symmetric interacting Hamiltonian in the dressed basis, namely
\begin{align}
\hat V_s =   \int d \vec{r}\left[  \frac{g_0}{2}\sum_{ij} \hat{\varphi}_{i}^{\dagger}\hat{\varphi}_{j}^{\dagger}\hat{\varphi}_{j}\hat{\varphi}_{i} + \tilde{g}_2 \left(\hat{\varphi}_{1}^{\dagger}\hat{\varphi}_{1} + \hat{\varphi}_{-1}^{\dagger}\hat{\varphi}_{-1}\right)\hat{\varphi}_{0}^{\dagger}\hat{\varphi}_{0} + \tilde{g}_2 \left( \hat{\varphi}_{1}^{\dagger}\hat{\varphi}_{-1}^{\dagger}\hat{\varphi}_{0}\hat{\varphi}_{0} + \hat{\varphi}_{1}\hat{\varphi}_{-1}\hat{\varphi}_{0}^{\dagger}\hat{\varphi}_{0}^{\dagger}  \right)\right],
\end{align}
with $\tilde{g}_2 = g_0\frac{\Omega^2}{16}\left(1 + O((\Lambda + \frac{\epsilon + \delta}{4})^2)\right)$. Proceeding analogously with the nonsymmetric part of the interaction potential, $\hat V_a = \frac{g_2}{2}\int d\vec{r}\sum_j (\hat{\vec{\psi}}^{\dagger} \hat{F}_j \hat{\vec{\psi}})^2 $,  yields corrections to Hamiltonian (2) in the main text of the order $g_2\Omega^2$, which are safely neglected since $\abs{g_2} \ll g_0$ for $^{87}$Rb.

\section{Mean-field simulations of the three-mode model}
\begin{figure}[b!]
\includegraphics[width=0.75\linewidth]{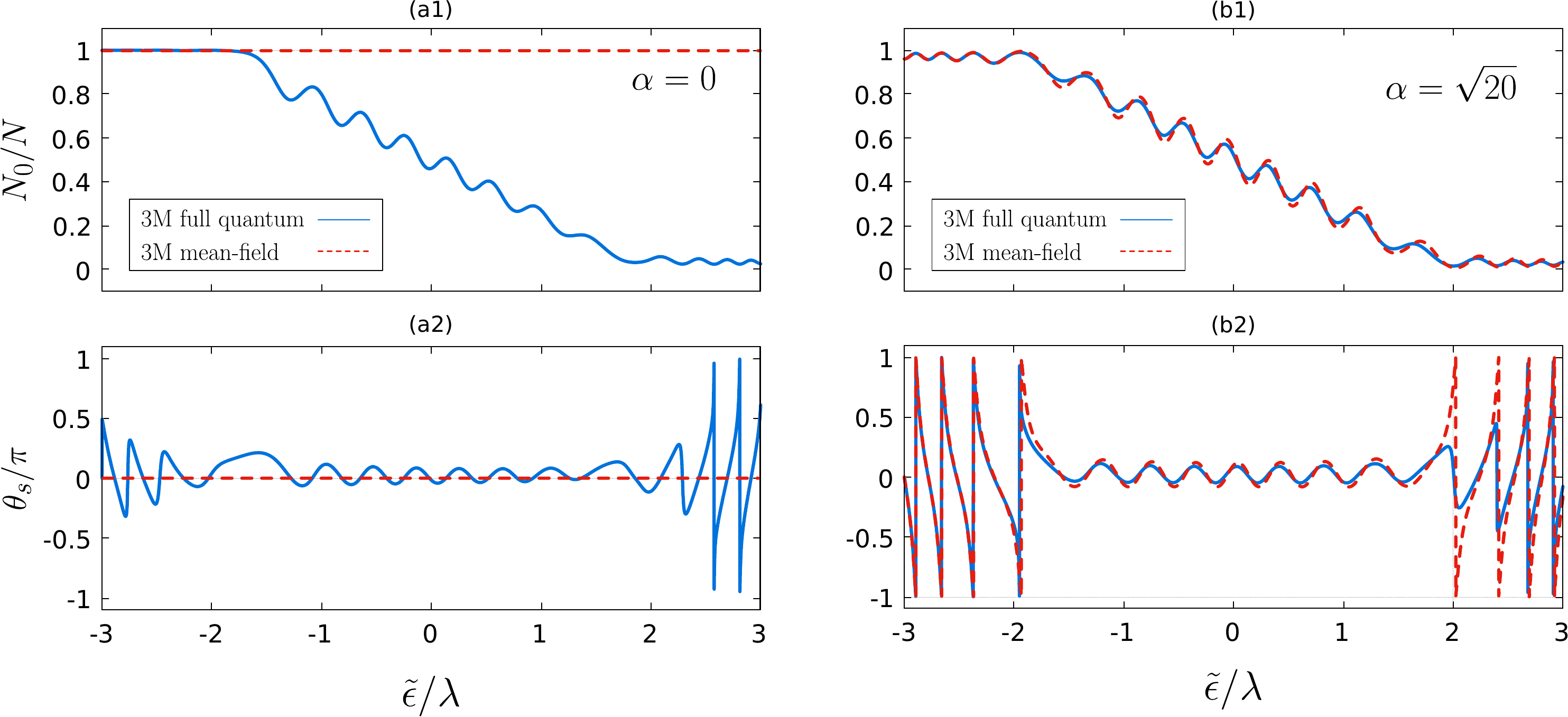}
\caption{(Color online) \textbf{Comparison between full quantum and mean-field simulations.} A state initially prepared at $\ket{N,\alpha,0}$ is driven from $\tilde{\epsilon} = -3\lambda$ to $\tilde{\epsilon} = 3\lambda$ with $\tau_d = 8h/\lambda$. The relative occupation of the state $\ket{\phi_0}$, $N_0/N$, along the drive is plotted in (a1) and (b1) for $\alpha = 0$ and $\alpha = \sqrt{20}$, respectively. In both cases $N=1000$. The corresponding spinor phase $\theta_s$ is plotted in (a2) and (b2). Blue solid lines show the results from full quantum simulations of Hamiltonian \eqref{3modeham}. Red dashed lines show the results obtained with the mean-field equations \eqref{3_mode_GPE}.}\label{Fig_3mode_simulations} 
\end{figure}

In the protocol described in the main text, the state approaches the Fock states $\frac{1}{\sqrt{N!}}(\hat{b}_0^\dagger)^N\ket{0}$ and $\frac{1}{(N/2)!}(\hat{b}_{-1}^\dagger)^{N/2}(\hat{b}_{1}^\dagger)^{N/2}\ket{0}$ while being in the P and TF phases, respectively. The mean field description of the evolution away from the BA phase is therefore expected to be inaccurate, with the dynamics being dominated by quantum fluctuations. Expressing Hamiltonian (4) in the main text explicitly in terms of the mode operators $\hat{b}_j$ yields
\begin{equation}\label{3modeham}
\ham_0 = \frac{\lambda}{N}\Bigg[(\hat{b}_{-1}^{\dagger}\hat{b}_1^{\dagger}\hat{b}_0 \hat{b}_0 + \textit{H.c.}) + \hat{N}_0(\hat{N}_1 + \hat{N}_{-1}) \bigg] - \tilde{\epsilon}\hat{N}_0.
\end{equation}
From eq. \eqref{3modeham}, the corresponding three-mode mean-field equations read
\begin{align}\label{3_mode_GPE}
i\hbar\dot{b}_1 &= \frac{\lambda}{N}\left[b_{-1}^* b_0 b_0 + b_0^* b_0 b_{1} \right], \nonumber \\
i\hbar\dot{b}_0 &= \frac{\lambda}{N}\left[2 b_{1} b_{-1} b_0^* + b_{1}^*b_{1} b_0 + b_{-1}^*b_{-1}b_0\right] - \tilde{\epsilon} b_0, \nonumber \\
i\hbar\dot{b}_{-1} &= \frac{\lambda}{N}\left[b_{1}^* b_0 b_0 + b_0^*b_0 b_{-1} \right],
\end{align} 
where we have identified $\mean{\hat{b}_{\pm 1,0}} = b_{\pm 1,0}$. Initially setting $b_{\pm 1} = 0$ or $b_{0} = 0$  into eqs. \eqref{3_mode_GPE} results in a stationary state, independently of $\tilde{\epsilon}$, in contradiction with the dynamics predicted by Hamiltonian \eqref{3modeham}. To address this issue, we test the effective model with the GPE of the full gas by simulating an analogous drive across the P-TF-BA excited diagram in a slightly lower lying family of excited states. As shown in \cite{Feldmann-prl-2021}, the properties of the excited phases of Hamiltonian (4) in the main text vary smoothly across the energy spectrum. Therefore, we instead prepare the initial state in a coherent state $\ket{N,\alpha,\theta_s} = \frac{1}{\sqrt{N!}}(\alpha\enum{-i\theta_s/2} \hat{b}_{-1}^\dagger + \sqrt{1-2\alpha^2}\hat{b}_0^\dagger + \alpha\enum{-i\theta_s/2} \hat{b}_1^\dagger )^N \ket{0}$, averaging $\alpha^2 > 0$ atoms in the pseudospin $\pm 1$ states. In these conditions, mean-field computations quickly converge to full quantum simulations as $\alpha$ is increased, as exemplified in Fig.\,\ref{Fig_3mode_simulations}, while the energy gap and the location of the phase boundaries do not vary significantly as long as $\alpha^2 \ll N$.

\section{Validity of the three-mode approximation}

\begin{figure}[b!]
\includegraphics[width=0.75\linewidth]{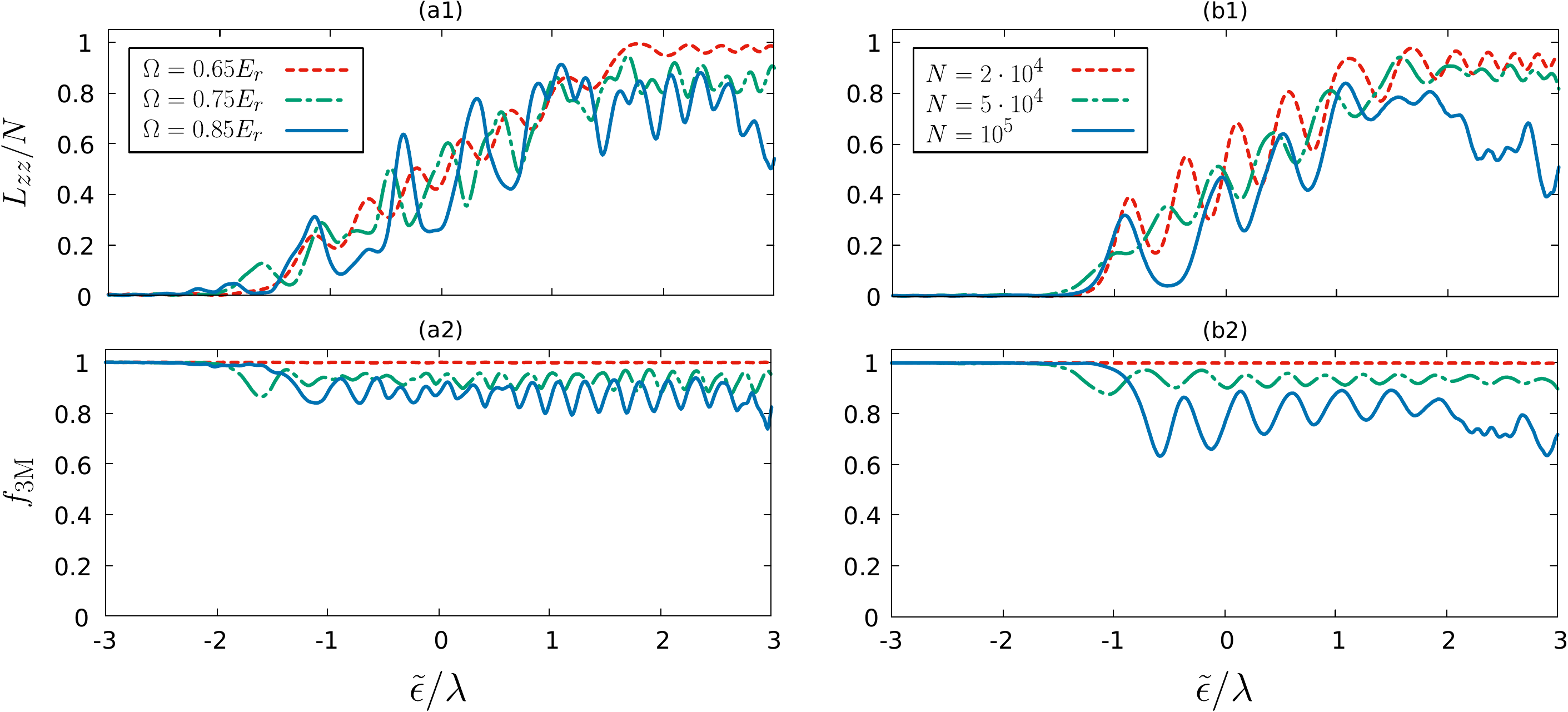}
\caption{(Color online) \textbf{Validity of the three-mode approximation.}  (a1) Expected value of $\hat{L}_{zz}$ as a function of $\tilde{\epsilon}$ for a state initially prepared at $\vec{\psi} = \sqrt{\alpha}(\vec{\phi}_{-1} + \vec{\phi}_{1}) + \sqrt{N-2\alpha}\vec{\phi}_{0}$, with $N=10^4$, $\alpha=25$ and $\hbar\omega_\mathrm{t}= 2\pi\cdot 140\,$Hz. The state is evolved under the GPE while driving $\tilde{\epsilon}$ from $-3\lambda$ to $3\lambda$, and keeping $\Omega = 0.65 E_r$ (dashed-red), $\Omega =0.75 E_r$ (dashed-dotted green) and $\Omega =0.85 E_r$ (solid blue). The total drive time is set to $\tau_d = 8h/\lambda$. (a2) Relative occupation of the three self-consistent modes $\vec{\phi}_{-1}$, $\vec{\phi}_{0}$ and $\vec{\phi}_{1}$, along the drive depicted in (a1).  (b1) Expected value of $\hat{L}_{zz}$ as a function of $\tilde{\epsilon}$ for a state initially prepared at $\vec{\psi} = \alpha(\vec{\phi}_{-1} + \vec{\phi}_{1}) + \sqrt{N-2\alpha^2}\vec{\phi}_{0}$, with $\alpha=\sqrt{10^{-3}N}$ and $N=2\cdot10^4$ (dashed-red), $N=5\cdot10^4$ (dashed-dotted green) and $N=8\cdot10^4$ (solid blue). The state is evolved under the GPE, driving $\tilde{\epsilon}$ from $-3\lambda$ to $3\lambda$, while keeping $\Omega = 0.5 E_r$ and adjusting $\omega_\mathrm{t}$ so that $n = 10^{-14}$\,cm$^{-3}$. The total drive time is set to $\tau_d = 8h/\lambda$. (b2) Corresponding relative occupation of the modes $\vec{\phi}_{-1}$, $\vec{\phi}_{0}$ and $\vec{\phi}_{1}$ along the drive depicted in (c1).}\label{Fig_few_mode_robustness} 
\end{figure}

The realization of stripe phases in an excited state permits to access the phase in regimes where it is experimentally more feasible. Notably, for nearly spin-symmetric BECs, the approach enhances the contrast of the spatial modulations in gas, which is proportional to $\Omega$. Since the protocol presented in the main text relies on the effective description of Hamiltonian (4) in the main text, we discuss here its validity. Hamiltonian (4) follows from a three-mode truncation of the Hilbert space, and it predicts the energy gap that is exploited in the quasi-adiabatic protocol to drive the state through a quantum phase transition. Qualitatively, the approximation is expected to be accurate for small condensates when $\abs{\lambda}, \abs{\tilde{\epsilon}} \ll g_0n, \hbar\omega_\mathrm{t}$. Nonetheless, it is difficult to quantitatively estimate its accuracy. To this end, we use the GPE of the full dressed and trapped spinor gas, and quantifies the accuracy of the approximation by computing the projection of the time-evolved states on to the subspace spanned by the three self-consistent mode, $f_\mathrm{3M} = \frac{1}{N^2}\sum_j \left\vert \int d\vec{r} \vec{\phi}_j^*\cdot\vec{\psi}\right\vert^2$,  previously computed via imaginary time evolution.

In the main text, we exemplify the realization of the protocol with simulated drives along two different trajectories in the $\Omega-\epsilon$ plane of the most excited phase diagram of the effective three-mode model. In both cases, the trajectories start at $\Omega = 0.65$, and we set $N(0) = 10^4$. In general, with $\tilde{g_2} \propto \Omega ^2$, the energy scale of the effective model is enhanced at larger $\Omega$, which reduces the preparation time and the relative impact of the heating mechanisms and of photon scattering loss. However, the validity of the three-mode model is progressively more challenged as $\Omega$ is increased. Similarly, at any given density, the robustness of the protocol strongly depends on the number of particles, as exemplified in Fig.\,\ref{Fig_few_mode_robustness}.

Note that different physical quantities are affected differently by the value of $f_\mathrm{3M}$. For instance, macroscopic entanglement preparation is expected to be very sensitive to the full quantum structure of the prepared state. Thus, even tiny reduction of $f_\mathrm{3M}$ are expected to result in considerable reduction of entanglement. On the contrary, the macroscopic spin transfer is less sensitive to the the leakage of probability amplitude out of the three-mode description, as exemplified in Fig.\,\ref{Fig_few_mode_robustness}. As a consequence, the optimal experimental parameters will be strongly dependent on the physical observables of interest.

\end{document}